\documentclass[trackchanges,twocolumn]{aastex7}

\usepackage{graphicx}
\usepackage{subfigure}
\usepackage{amsmath}
\usepackage{appendix}
\usepackage{hyperref}
\usepackage{multirow}
\usepackage[mathscr]{euscript}
\usepackage{xspace}

\newcommand{\Planck}{\textit{Planck}\xspace}

\newcommand{\ymap}{$y$-map\xspace}
\newcommand{\ymaps}{$y$-maps\xspace}

\defcitealias{KTully17}{KT17}
\defcitealias{Arnaud10}{A10}
\defcitealias{LeBrun15}{L15}

\bibpunct[; ]{(}{)}{,}{a}{}{;}

\graphicspath{{./}{figures/}}

\shorttitle{SZ-UNet}
\shortauthors{Pratt et al.}

\begin{document}

%\title{An Example Article using \aastex v6.2\footnote{Released on January, 8th, 2018}}
\title{Supervised Extraction of the Thermal Sunyaev\textendash Zel'dovich Effect with a Three-Dimensional Convolutional Neural Network}

\correspondingauthor{Cameron Pratt}
\email{campratt@umich.edu}

\author[orcid=0000-0002-6653-8490]{Cameron T. Pratt}
\affiliation{Department of Astronomy, University of Michigan, Ann Arbor, MI 48109, USA}
\email[]{campratt@umich.edu}

\author[orcid=0000-0002-2941-646X]{Zhijie Qu}
\affiliation{Department of Astronomy \& Astrophysics, The University of Chicago, 5640 S. Ellis Ave., Chicago, IL 60637, USA}
\affiliation{Department of Astronomy, Tsinghua University, Beijing, CN}
\email[]{quzhijie@tsinghua.edu.cn}

\author{Joel N. Bregman}
\affiliation{Department of Astronomy, University of Michigan, Ann Arbor, MI 48109, USA}
\email[]{jbregman@umich.edu}

\begin{abstract}
The thermal Sunyaev\textendash Zel'dovich (SZ) effect offers a unique probe of the hot and diffuse universe that could help close the missing baryon problem. Traditional extractions of the SZ effect, however, exhibit systematic noise that may lead to unreliable results. In this work, we provide an alternative solution using a three-dimensional Attention Nested U-Net trained end-to-end with supervised learning. Our labeled data consists of simulated SZ signals injected into \Planck frequency maps, allowing our model to learn how to extract SZ signals in the presence of realistic noise. We implement a curriculum learning scheme that gradually exposed the model to weaker SZ signals. The absence/presence of curriculum learning significantly impacted the amount of bias and variance present in the reconstructed SZ signal. The results from our method were comparable to those from the popular {\it{needlet internal linear combination}} (NILC) method when evaluated on simulated data as well as real-world SZ signals. We conclude by discussing future avenues for advancing machine learning extractions of SZ signals.

%We discuss that an optimal machine learning extraction method strongly depends on the expected signal-to-noise ratio, and this must be reflected in the loss function when using machine learning methods to extract the SZ signal.

\end{abstract}


\section{Introduction}
Over the past few decades, most of our knowledge about the hot, diffuse gas that permeates our Universe has come through X-ray observations. X-rays are easily observed for massive galaxy clusters, however, they are more challenging to detect with decreasing halo mass and/or increasing distance from the halo center. It is suspected that much of the baryons in the Universe are too diffuse to observe with current X-ray observatories, and detecting such gas would help resolve the ``missing baryon problem'' \citep{Bregman07,Shull12}. 

The thermal Sunyaev\textendash Zel'dovich (SZ) effect offers an independent view of hot and diffuse gas \citep{SZ1970,SZ1972,Birkinshaw97}. It is caused by the inverse Compton scattering of cosmic microwave background (CMB) photons with free electrons in a hot plasma. The amplitude of the SZ effect is given by the dimensionless parameter, $y$, which measures the integrated electron pressure along a line of sight
\begin{equation}
    y \propto \int_{0}^{\infty} n_{e}T_{e}dl
\end{equation}
where $n_{e}$ is the electron number density, $T_{e}$ is the electron temperature, and $dl$ is the differential path length. The SZ effect has a linear dependence on the density, making it a sensitive probe of low-density gas. On the other hand, Bremsstrahlung X-ray radiation has a quadratic dependence on the electron density, S$_{\mathrm{X}} \propto n_{e}^{2}$, making it less sensitive to diffuse environments. This is important in light of the missing baryon problem since most of the baryons are expected to exists as a diffuse medium. Therefore, the SZ effect could be an effective tool for observing the remainder of the missing baryons.

Extracting the SZ effect from observations is a challenging task. The expected signals are quite weak compared to other galactic and extragalactic components. It appears in from radio through infrared bands, causing a distortion to the CMB temperature on the order of $\frac{\Delta T}{T} \lesssim 10^{-4}$ which is for the strongest signals coming from the most massive clusters of galaxies \citep{Cavaliere91}. Real-world observations amount to a collection of photons coming from different physical sources, such as extragalactic dust emission and synchrotron radiation from background active galactic nuclei. In order to extract the SZ signal, one must solve an unmixing problem that separates the SZ component from all other signals.

A number of techniques have been developed to address component separation, especially for the CMB \citep[e.g.,][]{Bobin08,Leach08,Khatri19,Galloway23}. One of the best methods is known as the {\it{needlet internal linear combination}} \citep[NILC; ][]{Cardoso08, Delabrouille09, Remazeilles11}. NILC is a semi-blind extraction that aims to minimize the variance of the reconstructed SZ map (\ymap) while preserving the spectral dependence. Furthermore, it assumes zero correlation between the SZ effect and all other components which is empirically inappropriate. In turn, NILC is known to produce a level of systematic noise that contaminates the SZ extraction \citep{Pratt24}.

More recently, machine learning approaches with neural networks have migrated into this field in hopes of offering improvements. The continual progression of computing resources and datasets allow for training complex nonlinear models to capture the salient features within data. For example, \citet{Pratt24} implemented a machine learning pipeline to enhance the quality of NILC extractions. Similar concepts applied by \citet{McCarthy24} used a neural network to calculate a correction factor that was combined with an ILC solution to reduce noise levels in reconstructed \ymaps. Also, the authors of \citet{ResunetCMB22} developed a convolutional neural network (CNN) to extract the CMB polarization. Nevertheless, very few CMB/SZ studies have explored these techniques.

Machine learning models are often trained on labeled (ground truth) data via supervised learning. Unfortunately, perfectly labeled SZ signals do not exist in the real-world. In this work, we generate noisy labels using a set of mock SZ signals and injecting them into the latest frequency maps produced by the \Planck satellite \citep{PlanckDR4}. Then we perform end-to-end training to develop a model that can reliably extract SZ signals. 

The rest of the paper is structured in the following way: in \autoref{sec:methods} we describe the methodology; the main results are presented in \autoref{sec:results}; the results are discussed in \autoref{sec:discussion}; a summary is provided in \autoref{sec:summary}.

\section{Methods} \label{sec:methods}
The SZ effect is just one of many signals detected near the CMB frequencies ($\sim 23-857$ GHz). Other astrophysical (Galactic and extragalactic) and instrumental components must be separated to extract the SZ effect. In this section, we describe a machine learning pipeline built for this task. First, we create mock SZ observations and inject them into the \Planck frequency data. Then we provide details regarding our model architecture and training procedure. Finally, we briefly review the NILC algorithm so we compare its results with our developed model.

%/home/campratt/Dektop/RESEARCH/UDA/code/signal_addition_GL.py
%/home/campratt/Dektop/RESEARCH/UDA/code/plot_signal_background.ipynb
\begin{figure*}
    \centering
    \includegraphics[width=0.85\textwidth]{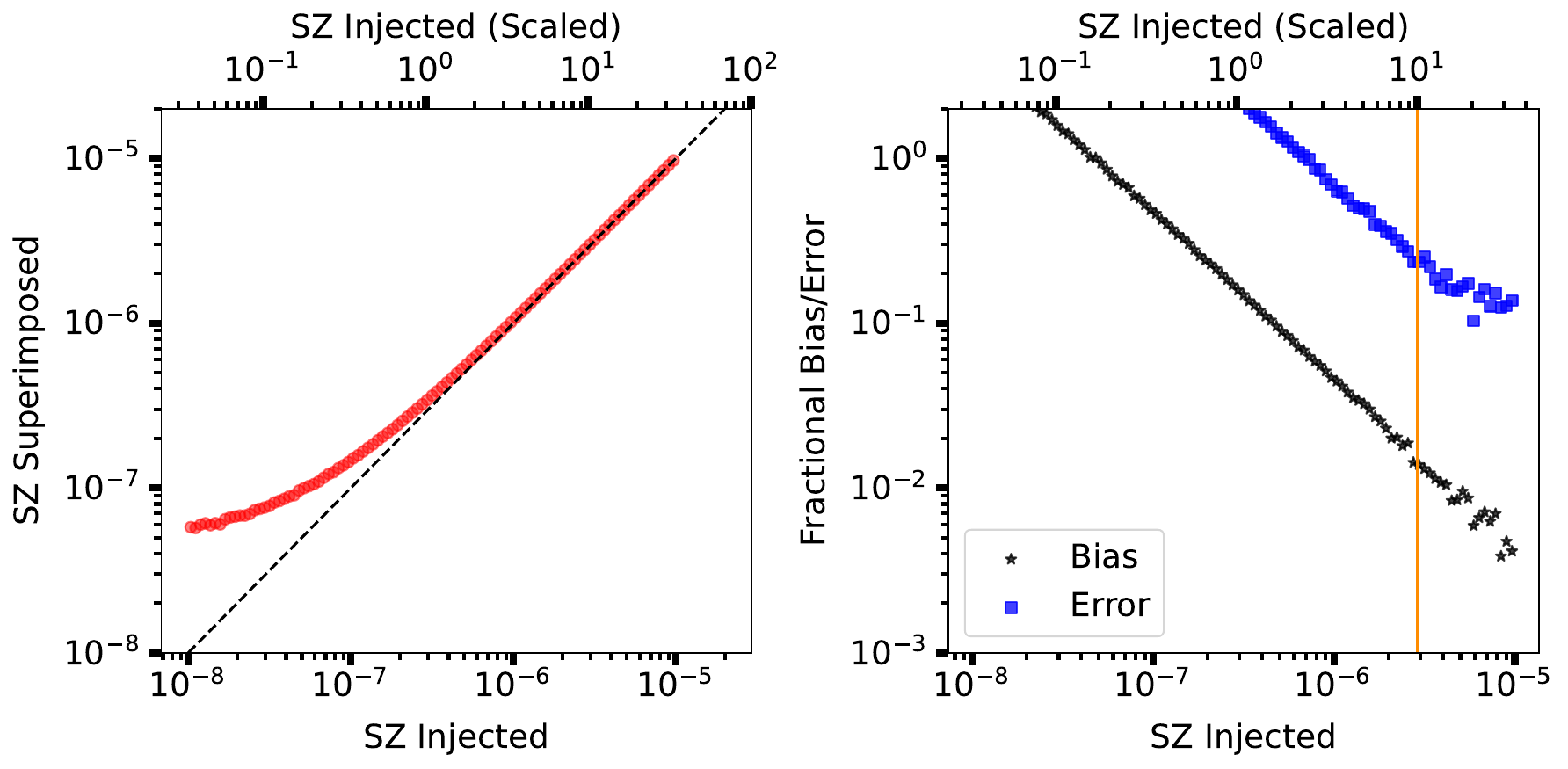}
    \caption{The effect of superimposing two SZ signals. (left) Averages of superimposed pixel values binned by the strength of the injected SZ signal. (right) Fractional biases and errors of the superimposed signal as a function of injected signal strength. In both panels, the upper x-axis shows the scaled SZ signal strength. The vertical orange line denotes were the scaled SZ signal equals 10.}
    \vspace{0.5cm}
    \label{fig:superimpose}
\end{figure*}
%%%%%%%%%%%%%%%%%%%%%%%%%%%%%%%%%%%%%%%%%%%%%%%%%%%%%%%%%%%%%%%%%

\subsection{Simulations} \label{sec:simulatoins}
The objective of this work was to train a model end-to-end with supervised learning to extract the SZ signal. Supervised learning requires a set of labeled SZ data, however, these do not exist in the real-world. This is because we only observe one universe, and it is this dataset that we want to use for prediction. Acquiring enough labeled SZ data for robust training required the use of simulations.

We chose to simulate SZ signals using the data products from \citet{Han21}. These authors constructed full-sky frequency maps based on the cosmological simulation runs from \citet{Sehgal10}. They generated a large suite of \ymaps by matching the spatial and spectral statistics from the full simulations. Similar to \citet{Pratt24}, \ymaps were derived from their 148 GHz maps using the known spectral dependence of the thermal SZ signal
\begin{equation}\label{eq:sz_temp}
    \frac{\Delta T}{T_{\mathrm{CMB}}} = g(\nu) y
\end{equation}
where $g(\nu)$ is the SZ spectral dependence, $\frac{\Delta T}{T_{\mathrm{CMB}}}$ is the change in temperature relative to the CMB. The synthetic \ymaps are publicly available in HEALPIX format \citep{Healpix} with an original resolution parameter of NSIDE$=4096$ which we downgraded to NSIDE$=2048$. 

We obtained one hundred full-sky \ymaps for our experiment. The synthetic \ymaps were then converted into SZ signals at each frequency using \autoref{eq:sz_temp}. The SZ frequency signals were then convolved by the \Planck channel beams and injected/superimposed into the fourth release of \Planck data \citep{PlanckPR4}. More discussion about the effects of superimposing SZ signals is provided in \autoref{sec:superposition}. The labeled \ymaps were smoothed to a final resolution of full-width at half-maximum of 10 arcminutes. Injected frequency maps served as model inputs while the smoothed \ymaps were the labeled outputs.

\label{sec:training}
\subsection{Train/Test Data}
The all-sky \ymaps and frequency maps were then modulated into square patches to work with traditional convolutional neural networks. Each input sample consisted of a $9\times 512 \times512$ multispectral cube; 512 pixels for the spatial dimension and 9 channels in the spectral dimension. The 2D patches were created using the Gnomonic projection in Healpy \citep{Healpy} with a resolution of 2 arcminutes pixel$^{-1}$. For each all-sky dataset, 700 multispectral cubes and SZ images were created. The center coordinates of the images were required to have Galactic latitudes ($b$) of $|b|>20^{\circ}$ and a minimum separation of $5^{\circ}$ from any other image center. This permitted partial overlap between images, but avoided creating identical patches for a single all-sky realization. This yielded 70,000 samples in total (700 images $\times$ 100 all-sky \ymaps), and the center positions of the 700 images were identical across all realizations. 

Data transformations were applied during the training process. First, we downsized the spatial dimension of each original sample to have 128 pixels per side rather than the full 512 to reduce the computational expense. During training, each reduced sample was taken from a random location within the original $512\times 512$ region. Random horizontal and vertical flipping of the images were also applied to assist with model generalization. Finally, we split the train/test data into 67:33 and applied three-fold cross-validation.

Input frequency data were normalized to zero mean and unit variance in the frequency dimension where the parameters $\mu$ and $\Sigma$ transform the original quantities $\tilde{z} = \frac{z - \mu}{\Sigma}$; these values were directly computed from PR4 data where $|b| > 20^{\circ}$, and are provided in \autoref{tab:freq_norms}. For the labeled SZ data, $\Sigma_{\mathrm{sz}}$ was calculated to be $2.895 \times 10^{-7}$ i.e., the standard deviation of all SZ pixels. We did not calculate $\mu$ for the output because the median of each SZ image was subtracted as a local background correction during training. 

Throughout this work, $\tilde{y}$ is defined as an SZ patch scaled by $\sigma_{\mathrm{sz}}$ and median subtracted. Working with $\tilde{y}$ was a necessary step because, otherwise, models would work too hard to get the background values correct. In this context, the background consists of the SZ signals coming from large-scale fluctuations. For example, contributions could arise from the Milky Way and the redshift-integrated background, where the former is expected to be much smaller than the latter. In fact, \citet{Chiang20} estimates the average cosmic SZ background to be $y \sim 1.2 \times 10^{-6}$ with some spatial variation depending on the distribution of large-scale structures. To this end, we decided to perform local background subtractions so the model does not have to focus on such variations; we are only interested in recovering the injected SZ signal strength rather than the absolute total.

Background subtractions, however, introduced uncertainty and the standard deviation of the background median values was $\approx 2.5\times10^{-8}$, which is about an order of magnitude lower than the spread of the input SZ values, $\Sigma_{\mathrm{sz}}$. Larger patches would result in even lower uncertainties, however the effect would be trivial. %However, the full range of these values was $\approx 3.5\times10^{-7}$.

\subsection{Superposition}
\label{sec:superposition}
As previously mentioned, the frequency data included the superposition of simulated SZ signals with the real sky. Doing this added power to the underlying, real-world SZ distribution which is largely unknown. In turn, the labeled data used in this work were considered noisy since they were added on top of a non-zero background.

First, we wanted to know the effect of injecting known SZ signals into a pre-existing background. Ideally, the superposition should be the real SZ $+$ simulated SZ, but we do not know the real SZ. To understand the effects of superimposing signals, we tested a simulated SZ $+$ simulated SZ superimposed dataset. This dataset was constructed by adding 10,000 random pairs of SZ patches from the full simulated sample. Each pair consisted of an ``injected'' and ``background'' signal. Both were assumed to be statistically representative of the underlying SZ distribution in the real Universe i.e., either image could represent the injected or background. The superimposed SZ signals were always larger than the injected signal. Median values were then subtracted from the superimposed map (injected $+$ background) and the injected map, converting it to the $\tilde{y}$ format used during training (see \autoref{sec:training}). 

The main effects of superimposing signals are displayed in \autoref{fig:superimpose}. The left panel shows the binned averages of the superimposed signals as a function of the injected signal strength. The right panel shows these same data but in terms of the average fractional bias $|\frac{\langle \tilde{y}_{2} - \tilde{y}_{1}\rangle}{\tilde{y}_{1}}|$ and average fractional error $\frac{\langle|\tilde{y}_{2} - \tilde{y}_{1}|\rangle}{\tilde{y}_{1}}$ where $\tilde{y}_{2}$ is the average superimposed SZ value for each bin of injected SZ signals $\tilde{y}_{1}$. 

\begin{figure}
    \centering
    \includegraphics[width=0.35\textwidth]{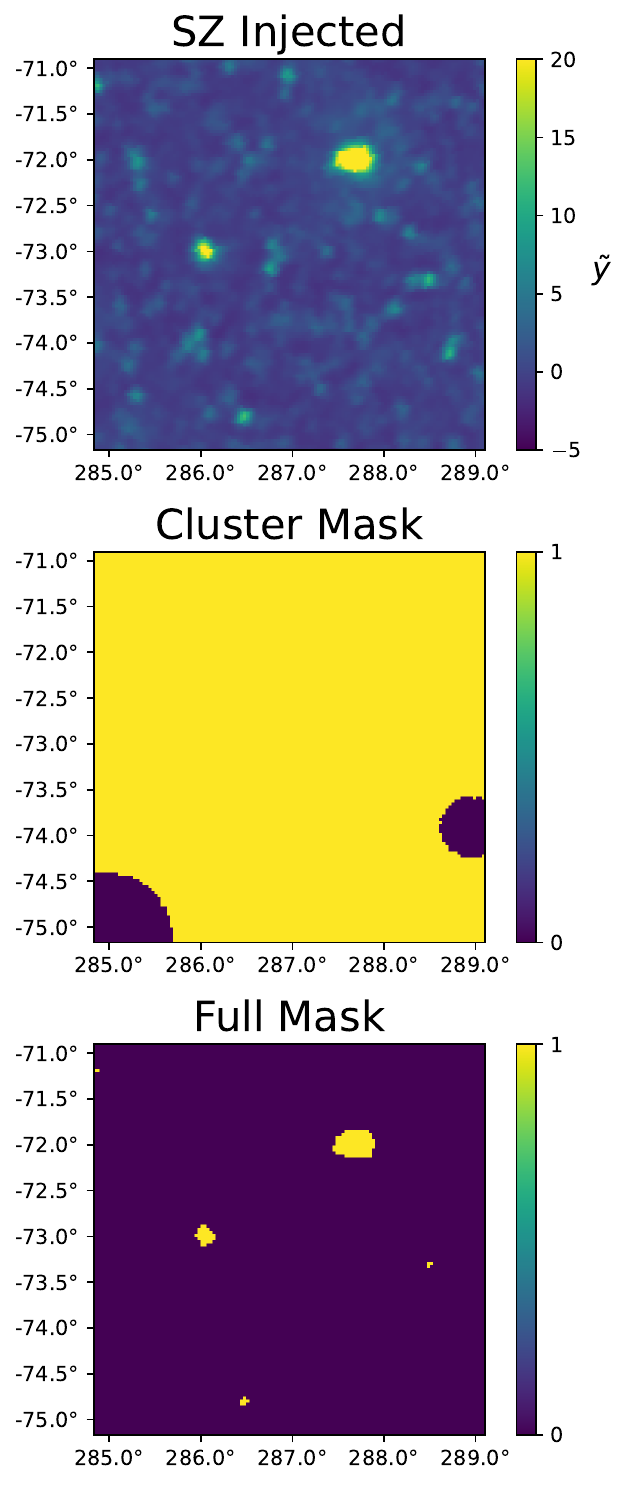}
    \caption{All cutouts are centered around the Galactic coordinates ($l,b$) = (286.97$^{\circ}$, -73.04$^{\circ}$) where the x-axis is Galactic latitude and y-axis Galactic longitude. (top) Panel of the SZ signals injected into the \Planck frequency maps. Notice that $\tilde{y}$ has negative values since the median pixel value was subtracted for a local background correction. (middle) A mask of known galaxy clusters provided by the PSZ2 and MCXC catalogs to prevent superimposing synthetic SZ sources on top of real signals. (bottom) A mask that includes the cluster mask from the middle panel as well as the threshold of $\tilde{y} > 10$ used for training.}
\label{fig:masking}
\end{figure}

An orange vertical line is drawn where the scaled SZ signal $\tilde{y} = 10$. This point denoted where the average fractional bias was $\approx 1\%$ and the average error $\approx 10\%$. We used this as an arbitrary lower limit where any values less than this threshold were considered to be inaccurate. In other words, this is where fluctuations from the background started to become too noisy relative to the injected signals. Only large enough SZ signals were used for training while smaller SZ signals were masked. This type of masking is shown as the bottom panel of \autoref{fig:masking}. 

Second, an additional subset of pixels were masked at the positions of known galaxy clusters. Real-world signals could lead to large uncertainties beyond the expected statistical fluctuations. We utilized the MCXC catalog \citep{MCXC} and the second catalog of \Planck SZ sources \citep[PSZ2;][]{PSZ2} to radially mask these objects out to $2 \mathrm{R}_{500}$ where R$_{500}$ represents the radius at which the average density is equal to 500 times the critical density. An example of this can be seen in the middle panel of \autoref{fig:masking}.

%%%%%%%%%%%%%%%%%%%%%%%%%%%%%%%%%%%%%%%%%%%%%%%%%%%%%%%%%%%%%%%%%

%\begin{figure}
%    \centering
%    \includegraphics[width=0.49\textwidth]{figures/ANUNet.png}
%    \caption{The ANU-Net architecture taken from Figure 4 of \citet{ANUNet}. Our network is modified so the double convolution block is three-dimensional, has a residual connection, and uses a leaky ReLU activation. Average pooling was used instead of maximum pooling, and only the spatial dimension was reduced in deeper layers. Outputs from the top layer were concatenated and then reduced to a single channel for the supervised task. All other aspects, including the attention gate module, were identical to the network used by \citet{ANUNet}.}
%    \vspace{0.5cm}
%    \label{fig:ANUNet}
%\end{figure}
%%%%%%%%%%%%%%%%%%%%%%%%%%%%%%%%%%%%%%%%%%%%%%%%%%%%%%%%%%%%%%%%%

\begin{table}[ht]
\centering
    \begin{tabular}{|c|c|c|}
        \hline
        \textbf{Frequency [GHz]} &
        \textbf{$\mu$ [K$_{\mathrm{CMB}}$]} &
        \textbf{$\Sigma$ [K$_{\mathrm{CMB}}$]} \\ \hline
        30 & $1.17097 \times 10^{-3}$ & $2.38019 \times 10^{-3}$ \\
        44 & $1.32803 \times 10^{-3}$ & $2.32772 \times 10^{-3}$\\
        70 &  $1.41115 \times 10^{-4}$ & $2.31159 \times 10^{-3}$  \\
        100  &  $3.07031 \times 10^{-5}$ & $ 2.32016 \times 10^{-3}$\\
        143  &  $5.86967 \times 10^{-5}$ & $ 2.33373 \times 10^{-3}$\\
        217  &  $3.01074 \times 10^{-4}$ & $ 2.40665 \times 10^{-3}$\\
        353  & $3.92069 \times 10^{-3}$ &$ 3.13457 \times 10^{-3}$ \\
        545  &  $5.62561 \times 10^{-2}$ & $ 3.60183 \times 10^{-2}$\\
        857 &  $4.06426$ & $2.67766$\\

        \hline
        
    \end{tabular}
    \caption{Mean and standard deviation values of the real frequency data that were used to normalize input data. Values were calculated using regions where $|b|>20^{\circ}$.}
    \label{tab:freq_norms}
\end{table}

%/home/campratt/Dektop/RESEARCH/UDA/code/plot_masks.ipynb

%/home/campratt/Dektop/RESEARCH/UDA/code/plot_validation_curve.ipynb
\begin{figure*}[t!]
     
    \centering
    \includegraphics[width=0.95\textwidth]{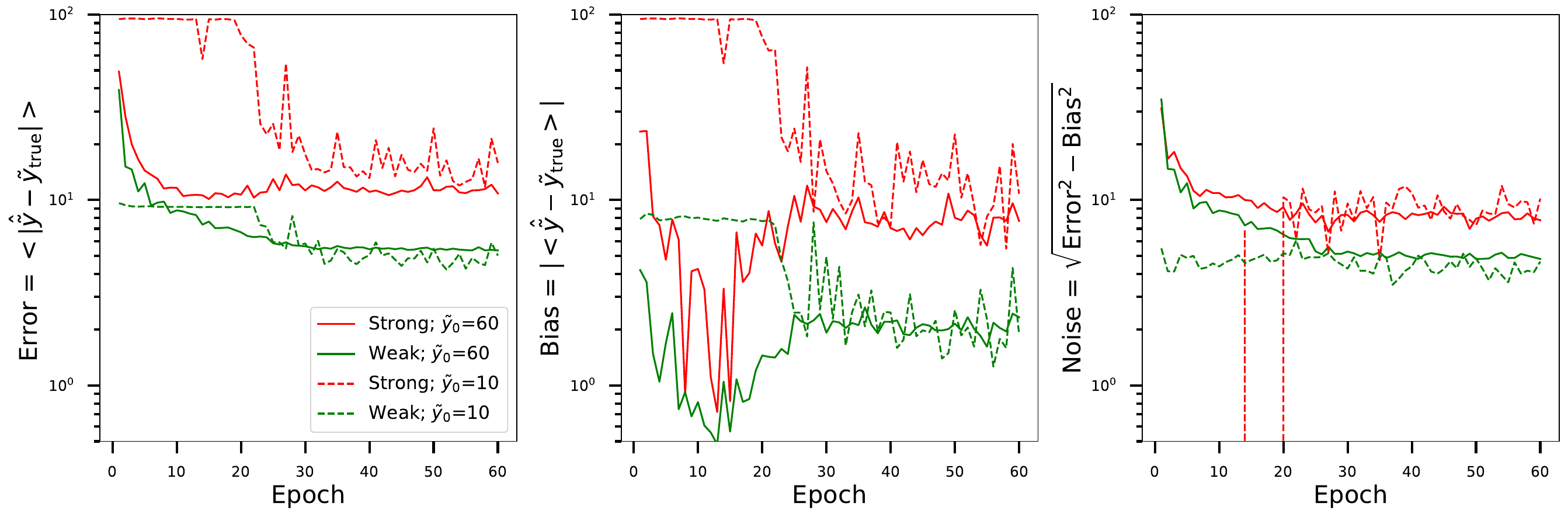}

    \caption{Errors, biases, and noise curves as a function of training epoch in the left, middle, and right panels respectively. The curves are the medians of the three cross-validation sets for the small model. Solid and dashed lines represent the results for $\tilde{y}_{0}$ values of 60 and 10 respectively. Red  curves monitor the performance of the strong pixels ($\tilde{y}>60$) and green for the weak signals ($10<\tilde{y}<30$).}
    
\label{fig:training}
\end{figure*}

\subsection{Modeling} 
\subsubsection{Architecture}
Here we present the network architecture and choice of hyperparameters. We sought an architecture that could both account for spatial and spectral changes in multispectral data, thus, capturing the SZ signal and foreground emissions. We selected the Attention Nested U-Net \citep[ANU-Net; ][]{ANUNet} as the backbone architecture and expanded it to three dimensions; we refer to it as SZU which is short for SZ-UNet. 

ANU-Net was built upon the original U-Net architecture, which consists of an encoder-decoder structure, and was originally created for medical imaging segmentation \citep{UNet}. The encoder generates a set of features at each layer where features from the previous layer are compressed in the spatial dimension by a factor of 2 while the channel dimension expands by a factor of 2. In the three-dimensional case, it is also possible to compress the spectral dimension, but we chose not to do this since the number of spectral channels was rather limited (only nine from the \Planck instruments). This process continues until the bottleneck. In a simple encoder-decoder architecture, the bottleneck contains all of the information needed to construct the desired output, however, spatial information becomes lost during the encoder stage. The U-Net famously introduced the skip connection so the decoder could use the spatial information previously gathered from the encoder for better results.

Over the past decade, modifications have been made to the original U-Net to advance its performance for a variety of tasks. One example being the Nested U-Net which introduced more skip connections between the encoder and decoder \citep{NestedUNet}. The idea was to further help the optimization process by bringing shallower information from the encoder to the deeper layers of the decoder. Additional progressions led to the development of ANU-Net which added attention gates. Attention gates act as filters during the skip connections by adding weights to the encoded features. These weights are guided by up-sampled features from deeper layers which helps the network determine which parts are relevant/irrelevant for the task at hand. Another benefit of ANU-Net is that the top layers have the same spatial dimensions as the inputs. Thus, both shallow and deep information can be leveraged in the loss function. This process is known as deep supervision which has shown to help facilitate the learning process \citep{Lee14,NestedUNet,Li22}.

The original network for 2D images is shown in Figure 4 of \cite{ANUNet}. Our network is modified so the double convolution block is three-dimensional, has a residual connection, and uses a leaky ReLU activation. Average pooling was used instead of maximum pooling, and only the spatial dimension was reduced in deeper layers. Outputs from the top layer were concatenated and then reduced to a single channel for deep supervision. All other aspects, including the attention gate module, were identical to the network used by \citet{ANUNet}.

\subsubsection{Loss Function}
The construction of the loss function was an important aspect in this work. In particular, we were forced to consider the unequal distribution of SZ signal strengths where weaker signals outnumbered stronger signals. A traditional loss function, such as the mean absolute error (MAE), would utilize all pixels, causing a bias toward the more abundant weak signals. In turn, less attention would be given to the stronger signals, leading to their poorer reconstruction. We attempted to alleviate this effect using curriculum learning.

Curriculum learning is a technique that facilitates training by first showing the model easy examples and gradually exposing it to more difficult ones. In our case, strong SZ signals were considered the easier examples. The loss function was designed as 
\begin{equation}
    \mathcal{L}(e) = \frac{1}{N_{\mathrm{sig}}}\sum_{i}^{N}{|\hat{\tilde{y_{i}}}} - \tilde{y_{i}}|; \tilde{y_{i}}>\tilde{y}_{\mathrm{sig}}(e)
\end{equation}
where $i$ indexes a single image, $\hat{\tilde{y_{i}}}$ is the predicted image, and $N$ represents the batch size. The subscript ``sig'' denotes a subset of pixels that exist above a certain value, $\tilde{y}_{\mathrm{sig}}$, with a total count of N$_{\mathrm{sig}}$. With curriculum learning, $\tilde{y}_{\mathrm{sig}}$ changes with each training epoch, $e$, and decreases by a value of $\alpha=2$ as
\begin{equation}
    \tilde{y}_{\mathrm{sig}}(e) = \tilde{y}_{0} - \alpha e
\end{equation}.
All pixel values of $\tilde{y}<\tilde{y}_{\mathrm{\mathrm{sig}}}$ were masked from the loss function during a given training epoch. A lower limit was set for $\tilde{y}_{sig}=10$ given the aforementioned restrictions arising from superimposing signals. Pixel values below this limit were always masked during training. An example of the pixel masking strategy outlined above is shown in \autoref{fig:masking}. As $y_{\mathrm{sig}}$ decreased each epoch, the number of pixels included in the loss function increased. The value of $\tilde{y}_{sig}$ at the zeroth epoch, $\tilde{y}_{0}$, was set arbitrarily to values of 60 and 10; these denote models with and without curriculum learning. 

\subsubsection{Model Selection}
In order to select a model for analysis, we considered the performance of our model for a range of SZ values: strong $\tilde{y}_{\mathrm{strong}} > 60$, intermediate $60>\tilde{y}_{\mathrm{int}}>30$, and weak $30>\tilde{y}_{\mathrm{weak}}>10$ signals. All models were trained for 60 epochs, and the performances at every epoch were considered during model selection. Ideally, we wanted a model with low bias ($|\langle \hat{\tilde{y}} - \tilde{y}_{\mathrm{true}}\rangle|$) and total error ($\langle |\hat{\tilde{y}} - \tilde{y}_{\mathrm{true}}|\rangle$) for all signal strengths, where $\tilde{y}_{\mathrm{true}}$ is the injected signal. For simplicity, we rank-ordered the errors and biases over all epochs for the three classes of signal strengths. The best model was chosen as the one with the lowest sum of ranks. We acknowledge the arbitrariness of our selection criteria. Devising such a method was not straight forward because because the best model depends on the specific scientific task. We discuss this further in \autoref{sec:BVT} in context of the bias-variance trade-off. 

%/home/campratt/Dektop/RESEARCH/UDA/code/plot_validation_curve.ipynb
\begin{table*}[t!]
    \centering
    \begin{tabular}{|c|cc|cc|}
        \hline
        \multirow{2}{*}{} & \multicolumn{2}{c|}{Small Model} & \multicolumn{2}{c|}{Large Model} \\
        \cline{2-5}
        & $\tilde{y_{0}}$=10 & $\tilde{y_{0}}$=60 & $\tilde{y_{0}}$=10 & $\tilde{y_{0}}$=60 \\
        \hline
Weak Bias & $\mathbf{-0.66^{+0.19}_{-1.73}}$ & $-1.62^{+1.56}_{-0.25}$ & $\mathbf{-0.50^{+0.23}_{-0.25}}$ & $-0.78^{+0.37}_{-0.92}$ \\
Intermediate Bias & $-5.11^{+0.66}_{-0.99}$ & $\mathbf{-2.28^{+2.53}_{-0.82}}$ & $-4.15^{+0.04}_{-0.78}$ & $\mathbf{-1.08^{+0.52}_{-1.16}}$ \\
Strong Bias & $-6.94^{+1.49}_{-0.58}$ & $\mathbf{-5.20^{+4.38}_{-0.92}}$ & $-5.08^{+1.06}_{-0.58}$ & $\mathbf{-4.40^{+3.15}_{-0.22}}$ \\
Weak Error & $\mathbf{4.51^{+0.91}_{-0.94}}$ & $5.67^{+2.49}_{-0.17}$ & $\mathbf{3.94^{+0.48}_{-0.02}}$ & $6.68^{+0.39}_{-1.14}$ \\
Intermediate Error & $8.03^{+1.68}_{-0.04}$ & $\mathbf{6.71^{+1.57}_{-0.13}}$ & $7.72^{+0.99}_{-0.08}$ & $\mathbf{7.13^{+0.17}_{-0.52}}$ \\
Strong Error & $11.77^{+4.16}_{-0.17}$ & $\mathbf{10.04^{+0.54}_{-0.09}}$ & $10.68^{+1.81}_{-0.80}$ & $\mathbf{9.80^{+0.31}_{-0.51}}$ \\
\hline
    \end{tabular}
    \caption{Validation metrics for different signal strengths given the four tested models. The values represent the medians while error bars represent the bounds obtained from the three cross-validation sets. Bold values show the smallest reconstructed error/bias for the small and large models. Curriculum learning ($\tilde{y}_{0}$=60) yielded the most unbiased solutions. The results from the large and small models were indistinguishable given the spread in cross-validation errors.}
    \label{tab:best_models}
\end{table*}

\subsubsection{Implementation Details}
Machine learning models require a number of hyperparameters to be fixed or tuned throughout the training process. An important value was the batch size due to the pixel masking scheme described above. Selecting a batch size too small could return empty datasets during early epochs when the pixel masking was most prominent and would be inoperable for training. A batch size of 128 was found to be sufficient, and increasing this value had no significant impact on the results. We also tested two different sets for the number of nodes used at each layer: these were deemed the small and large models. The number of nodes at each layer (from shallowest to deepest) were [8,16,32,64,128] and [16,32,64,128,256] respectively. 

We used the Adam optimizer with an initial learning rate of $10^{-3}$ and (.99, 0.999) as the first and second moment parameters in PyTorch. The value of $\alpha=2$ was fixed for all models. Gradients were also normalized to unity to stabilize training. All other training parameters were set to the default values in Pytorch. Training was executed out to 60 epochs and the learning rate was reduced by a factor of two at 30 epochs. The small models were trained on two NVIDIA A40 GPUs using data parallel multiprocessing for roughly four days; the large models used four GPUs and trained in approximately two days.

%No significant differences were apparent between model sizes. However, more exhaustive efforts should be made in the future to understand the effects of model structure, both in size of the Nested 3D U-Net as well as other exploring other architectures.

\subsection{Evaluation of the Real Sky}
The goal of training a model with simulated data was to eventually use it for real-world predictions. Then we wanted to compare the results with other extraction techniques. We chose to compare with the well-known NILC method using the same input frequency maps. 

%First, we define metrics that compare performances between the SZU and NILC algorithms using simulated data. Then we analyze the results on a subset of SZ objects in the second catalog of \Planck SZ sources \citep[PSZ2;][]{PSZ2}. 

\subsubsection{NILC}
NILC is an analytical, semi-blind component extraction technique. It is semi-blind in the sense that it only relies upon the theoretical spectral dependence of the SZ signal. Its closed analytical form makes it easy to implement, and it relies on few assumptions. The main assumptions are the definition of two hyperparameters: a wavelet basis and a spatial window function. In this work, we follow a similar approach to \citet{Pratt24}, specifically using their spatial parameter $\Gamma=10$ which yielded unbiased solutions. However, we used a slightly different wavelet basis, $\psi = \{10'-20',20'-40',40'-80',80'-160',160'-320',320'-640',640'\}$, but the specific choice of wavelet basis is trivial compared to the selection of $\Gamma$ \citep{Pratt24}. Additionally, we did not include data from the Wilkinson Microwave Anisotropy Probe that were included by \citet{Pratt24}. They were omitted here for consistency with SZU which utilized only the nine \Planck channels primarily for light-weight calculations.

The authors of \citet{Pratt24} closely followed the original NILC framework presented by \citet{Delabrouille09} which we refer the reader for a more comprehensive definition. In this work, we followed the same general principles with a few minor variations. First,  

In this work, we did not split the extraction into {\it{analysis}} and {\it{synthesis}} steps. In the original implementation, half of the wavelet transformation ($\psi^{0.5}$) was applied in harmonic space before the component extraction (analysis), and then applied again afterwards (synthesis). Instead, we applied to full wavelet transform once, mainly to handle the different instrumental beams by convolving the frequency maps to a common resolution. For example, all frequency maps with resolutions higher than 10$^{\prime}$ (i.e., from the \Planck high-frequency instruments) were smoothed to a common resolution of 10$^{\prime}$ and 20$^{\prime}$ respectively in order to apply the first of the seven wavelet transformations. 

Second, the smoothing was performed in the pixel domain rather than using spherical harmonics. Working in harmonic space can introduce high frequency noise in the presence of strong point sources. Conventionally, point sources are masked and in-painted before applying the harmonic transformation to avert this issue. However, the same issue can be avoided by smoothing in pixel space at the cost of slightly more expensive calculations. In general, the differences between the original process and the one used here were negligible.  

%%%%%%%%%%%%%%%%%%%%%%%%%%%%%%%%%%%%%%%%%%%%%%%%%%%%%
%/home/campratt/Dektop/RESEARCH/UDA/code/plot_analyze_evaluate_sz.ipynb
\begin{figure*}[t!]
    \centering
    \includegraphics[width=0.98\textwidth]{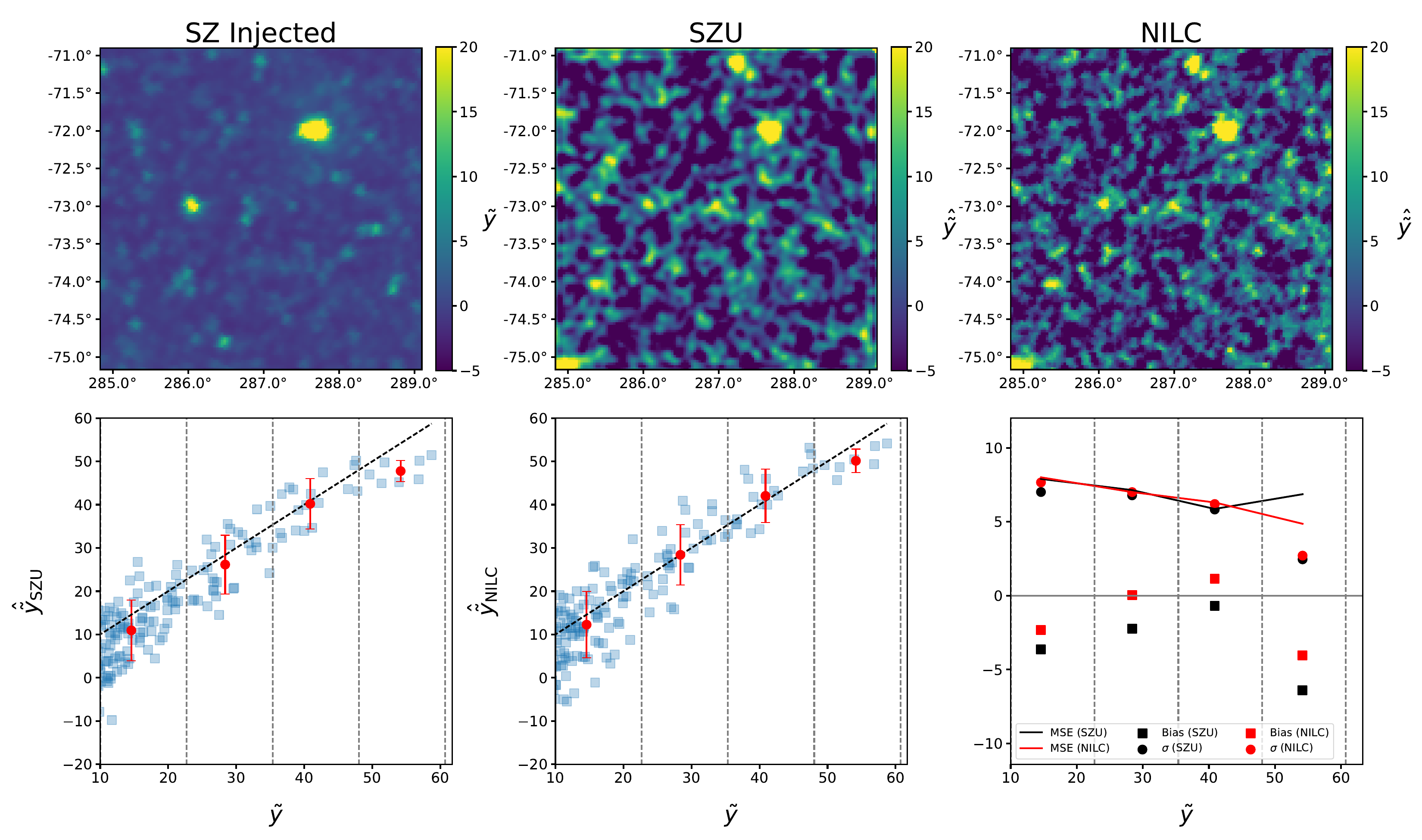}
    \caption{Example of extractions for a patch of sky centered around the same coordinates in \autoref{fig:masking}. The top row shows visual maps of the input SZ signal, the recovered SZU map, and the NILC extraction from left to right. The bottom-left and bottom-middle panels show the pixel values of SZU and NILC in seven linearly spaced bins (dotted vertical gray lines) and a 1:1 line as a dashed black line. The bottom-right panel shows the bias, scatter, and MSE values for SZU and NILC in black and red respectively. Both recover a strong SZ source in lower left corner which is a known galaxy cluster as shown in \autoref{fig:masking}.}
    \label{fig:example_plot}
\end{figure*}

\subsubsection{PSZ2 Clusters}
In order to get a sense of performance between SZU and NILC, we utilized the PSZ2 catalog which consists of over 1600 sources. We reduced this to a smaller subsample during analysis to conform to the caveats of our method. The first restriction was to limit the sample to clusters where $\theta_{500} = \frac{R_{500}}{D} < 20'$, where $D$ represents the angular diameter distance, such that we could extract the SZ flux out to 5$\theta_{500}$. This restriction was placed because we worked with relatively small sky patches $\sim 4^{\circ}\times 4^{\circ}$ to train SZU. It excluded extended systems such as the Coma and Virgo clusters which span many degrees on the sky \citep{PlanckComa,PlanckVirgo}. Additionally, this retained objects with estimated values for the mass enclosed with R$_{500}$ and redshift, $z$, since not every entry in the original catalog contained these estimates, and both were needed to obtain $\theta_{500}$. The second requirement was to place a lower limit of absolute Galactic latitude $|b|>20^{\circ}$, which was also implemented in training SZU. 

The PSZ2 catalog was constructed using a matched multi-filter technique \citep{Melin06} assuming a radial cluster profile shape defined by the universal pressure profile
\begin{equation}
    \frac{P(r)}{P_{500}} = \frac{P_{0}}{x^{\gamma}(1+x^{\alpha})^{\frac{\beta-\gamma}{\alpha}}}
\end{equation}\label{eq:UPP}
where $x=\frac{r}{R_{500}}$, the values of $P_{500}$, $P_{0}$, $\gamma$, $\alpha$, and $\beta$ were determined using the values from \citet{Arnaud10}. We utilized this same profile shape along with the values of the enclosed mass within R$_{500}$, M$_{500}$, and $z$ from the catalog to define the theoretical angular SZ profile $y_{upp}(\theta | M_{500}, z)$.

For each object, we measured their SZ radial profile by taking averages within 10$^{\prime}$ radial bins out to two degrees. Then we fit the data with a function that convolved $y_{upp}$ with a circular Gaussian beam of full-width at half-maximum of 10', which corresponds to the highest resolution defined by the NILC wavelet basis and SZU labels. A normalization factor was added as a free parameter into the fitting routine for a better representation of the radial profile, and we call this new profile $y_{\mathrm{fit}}$. 

After fitting the radial profiles, we extracted the SZ ``flux'', $Y_{\mathrm{SZ}}$, from sources in the PSZ2 catalog out to 5$\theta_{500}$
\begin{equation}
    Y_{\mathrm{SZ}} = 2\pi \int_{0}^{5\theta_{500}} y_{\mathrm{fit}}(\theta) \theta d\theta
\end{equation}. This is an observable that, when combined with a distance measurement, is related to the total thermal energy contained within a system. Making an assumption about the temperature then leads to an estimate of the gas mass which is a critical cosmological measurement.

\section{Results}\label{sec:results}
In this section, we present the main results. First, we describe the results of training SZU, focusing on the validation metrics. Second, we perform a comparison between our trained SZU model and the popular NILC method using simulated data and the PSZ2 catalog.

\subsection{Validation Metrics}
%%%%%%%%%%%%%%%%%%%%%%%%%%%%%%%%%%%%%%%%%%%%%%%%%%%%%%%%%%%%%%%%%

We tracked the average error and bias over 60 epochs for three cross-validation sets. The medians of the three cross-validation sets for the small models are shown in \autoref{fig:training}. It highlights how the validation metrics generally changed as training progressed depending on the starting values of $\tilde{y_{0}} = 60, 10$. We omitted the values for the intermediate signal range in this plot to avoid over-crowding. 

The model with no curriculum learning ($\tilde{y_{0}}=10$) appeared to struggle learning during early epochs which is demonstrated by the horizontal dashed curves in the both panels. It appeared to take $\sim 20$ epochs for the bias and error to begin decreasing. The model with curriculum learning ($\tilde{y_{0}}=60$), however, started learning immediately and eventually achieved better overall performance for the strong signals.

In the bias plot, the models with curriculum learning (solid lines) achieved smaller biases for the strong signals. In the error plot, the performance on the strong signals (red) was also better. For the weak signals (green), however, models with and without curriculum learning eventually achieved similar bias levels. The absence of curriculum learning also yielded slightly better performance in the total error and noise of the weak signals. 

The best models for all $\tilde{y_{0}}$ values, signal ranges, and model sizes are provided in \autoref{tab:best_models}. The superscript and subscript values determine the spread from the three-fold cross-validation. For each model size, we determined which $\tilde{y}_{0}$ yielded the smallest absolute bias and error shown in bold. The trends were similar between the small and large models as well as their global performances. It suggests model size was not an important factor since all values were roughly within the cross-validation uncertainties.

\subsection{SZU vs. NILC}
Here we show the performances of SZU and NILC. The SZU model used the $\tilde{y}_{0}=60$ model selected at the epoch where the rank-ordering scheme was a minimum. We compared the two methods using a particular set of three validation datasets taken from one of the cross-validations. For simplicity, we call this the comparison set. The comparison set consisted of 2100 sky patches since 700 patches make one full-sky set. We could have used more than three full-sky datasets, however, NILC was computationally expensive. 

%\subsubsection{Pixel Value Comparison}

In \autoref{fig:example_plot} we provide a visual example of SZU and NILC for a random patch of simulated sky (the same location provided in \autoref{fig:masking}). The top panels show the reconstructed images using SZU and NILC as well as the true input. A comparison of the binned pixel values are shown in the bottom panels. The bias, scatter/standard deviation ($\sigma$), and their mean squared error (MSE) of the binned data are shown in the bottom right.

\begin{figure}[t!]
    \centering
    \includegraphics[width=0.44\textwidth]{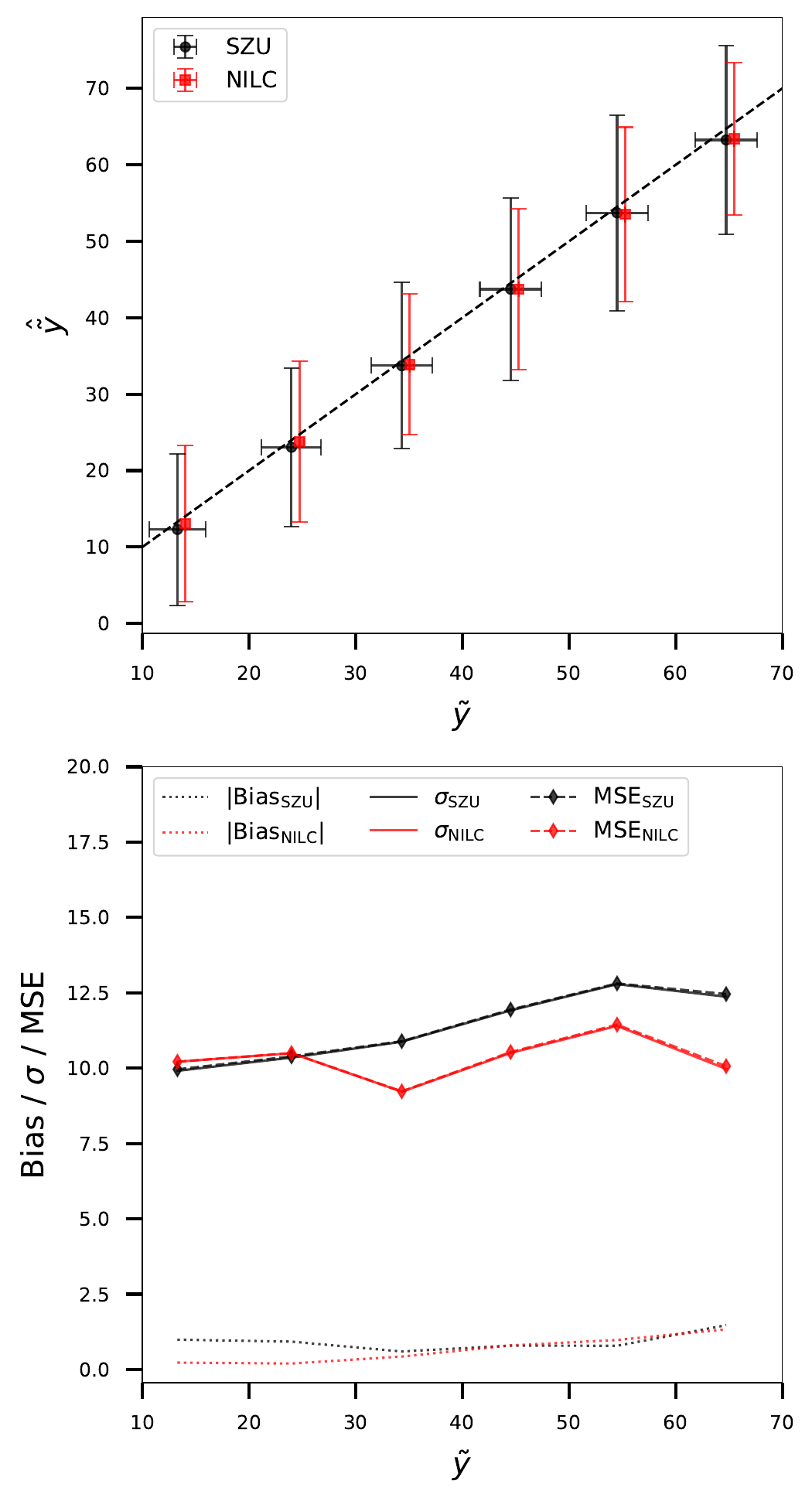}
    \caption{(top) Recovered pixel values for SZU (black) and NILC (red) for the comparison sample. Data are binned by the true values ($\tilde{y}$) and red data are slightly offset to the right for visual purposes. (bottom) Statistics of the biases (dotted), standard deviations (solid), and MSEs (dashed with diamonds). SZU exhibits more bias for the smallest $\tilde{y}$ values and more MSE for the larger $\tilde{y}$ values.} %A few known galaxy clusters are present in this field which were identified in all-sky X-ray and SZ catalogs \citep{MCXC,PSZ2} at pixel positions near (x=60,y=100) and (x=45,y=50).}
    \label{fig:pixel_compare}
\end{figure}

Similar data are also displayed in \autoref{fig:pixel_compare}. In this plot, however, we show the statistics for the full comparison sample. In the top panel, we show the estimated pixel averages against the input averages; the biases, scatters, and MSEs for the binned pixel values are shown in the bottom plot. Both panels illustrate similar performances between SZU and NILC. The MSEs of SZU and NILC were dominated by the statistical scatter which is why the dashed lines are obfuscated by the solid lines. SZU appeared to have more bias but less MSE toward weaker signals compared to NILC. For larger signals, the biases of SZU and NILC were more similar, but SZU exhibited a larger MSE. The differences between the MSEs of NILC and SZU were on the order of a few percent for all values of $\tilde{y}$.

%/home/campratt/Dektop/RESEARCH/UDA/code/plot_pixel_comparison.ipynb
%%%%%%%%%%%%%%%%%%%%%%%%%%%%%%%%%%%%%%%%%%%%%%%%%%%%%%%%%%%

\subsubsection{PSZ2 Comparison}
\begin{figure*}[t!]
    \centering
    \includegraphics[width=0.95\textwidth]{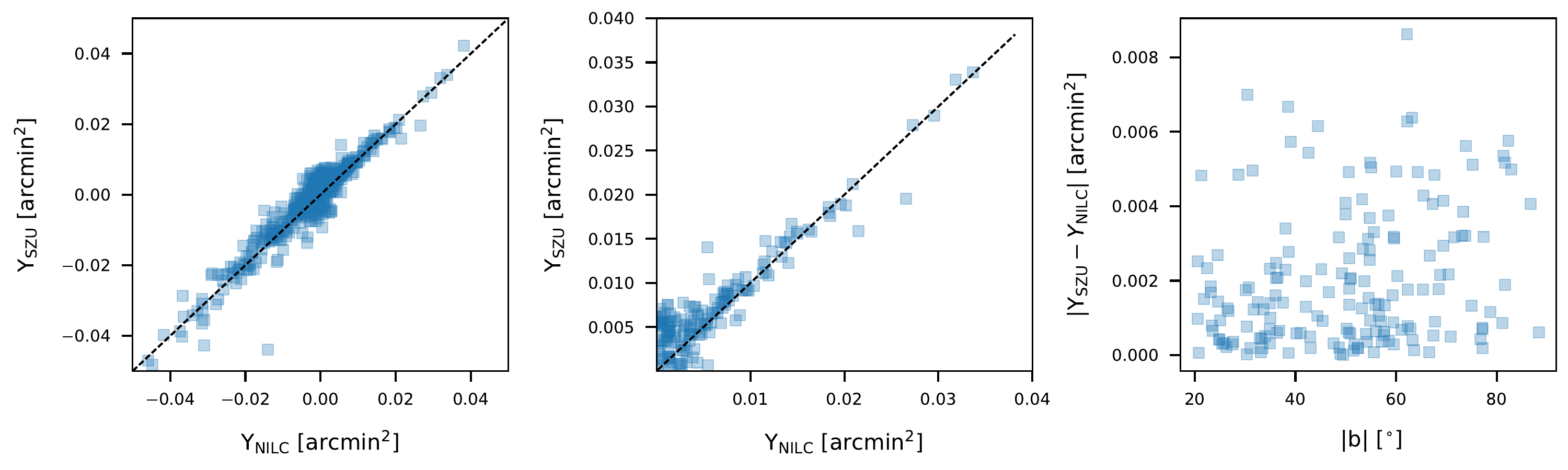}
    \caption{Comparisons of extracted Y$_{\mathrm{SZ}}$ values for SZU and NILC using the PSZ2 catalog. The left panel shows the relation between the two methods for the full sample while the middle panel provides a zoomed-in version for the positive signals. The dashed black line in these plots shows the 1:1 line. The right panel shows the absolute differences as a function of absolute Galactic latitude for the subsample given in the middle plot.} 
    \label{fig:YSZ_compare}
\end{figure*}

Both SZU and NILC were evaluated on real-world examples identified in the PSZ2 catalog. We extracted Y$_{\mathrm{SZ}}$ by integrating their fitted radial profiles out to 5$\theta_{500}$ and plotted the relation between NILC and SZU in \autoref{fig:YSZ_compare}. The left panel shows the values of the full sample while the middle panel includes only positive points. Negative SZ values are not physically possible, however, they may arise due to statistical fluctuations near the background level because the median was subtracted. Strong negative signals in SZU and NILC could be due to the presence of foreground contamination. This may occur when intense foreground contamination is correlated with the SZ spectral dimension, leading to an over-correction. This is commonly seen in NILC when there are strong radio signals emanating from clusters that host active galactic nuclei. The MMF methods used in PSZ2 also have systematic uncertainties, and many of the objects in the PSZ2 catalog are likely false positive signals. 

The right panel shows the absolute differences between NILC and SZU as a function of absolute Galactic latitude. The residuals do not exhibit any correlation, suggesting NILC and SZU handled foregrounds in a similar manner.

%/home/campratt/Dektop/RESEARCH/UDA/code/plot_psz2_ML_nilc.ipynb
%%%%%%%%%%%%%%%%%%%%%%%%%%%%%%%%%%%%%%%%%%%%%%%%%%%%%%%%%%%

\section{Discussion} \label{sec:discussion}

This work trained a machine learning model, called SZU, to extract the SZ signal with supervised learning. Here we discuss the results presented above and offer suggestions for future improvements.

\subsection{Bias and Variance}\label{sec:BVT}
%\subsubsection{Training Metrics}
The results presented in \autoref{fig:training} highlight the differences in model performance with curriculum learning. Curriculum learning had a significant effect in reducing the bias for intermediate and strong signals. It also facilitated the training process by helping the model learn the strongest and sparsest SZ signals then gradually learn how to handle the weaker and more abundant signals. Training with low no curriculum learning, however, struggled to capture the strong signals during early epochs and yielded a significant source of bias.

%\subsubsection{Selected Models}
Curriculum learning yielded the best overall models according to the selection criteria. This is reflected in \autoref{tab:best_models} where the best performances (smallest absolute values) for each metric are in bold. These results suggest that curriculum learning should be preferred for an unbiased solution for the intermediate and strong signals; the bias for the weak signals were comparable with or without curriculum learning. No curriculum learning yielded less total error for the weaker signals, albeit by a very small amount.   

Building on that point, selecting which model to use depends on the scientific task at hand. If one is investigating strong SZ signals, then curriculum learning would be preferred since those models yielded the lowest errors and biases for this regime. For weak signals, however, the total overall error was smaller without curriculum learning. 

Moreover, choosing which model to use becomes complicated when stacking a large sample of weak signals. The ultimate goal should be to maximize the signal-to-noise ratio
\begin{equation}
    \mathrm{SNR} = \frac{y + B}{\sqrt{Var(\hat{y})^{2} + B^{2}}}
\end{equation}
where, $\hat{y}$ is the estimated signal, $y$ is the true underlying signal, $B$, is bias, and $Var(\hat{y})$ represents the statistical noise. Ideally, we want to minimize the total contribution from the denominator while maintaining a strong signal in the numerator. Stacking reduces the statistical noise by a factor of $\sqrt{N}$, where $N$ represents the size of the stacking sample, under the assumption of symmetric Gaussian noise. If the goal is to stack many SZ signals to where $Var(\hat{y})^{2}<<B^{2}$, then a using a low-bias extraction would be imperative to return accurate results.

It is important to note that what constitutes a ``better'' SZ extraction is subjective. If the goal is to investigate strong signals or perform stacking, then an unbiased extraction should be preferred. On the other hand, if the goal is to study weak individual signals, such as the outskirts of a galaxy cluster, then a small amount of bias may be tolerated in exchange for lower statistical noise. In the future, devising an optimal extraction method will need to be tailored to the specific object(s) of study. The anticipated signal strength and noise level will have to be estimated beforehand.

Lastly, we emphasize that our model selection used a rank-based method to balance the total error and bias from three classes of signals strengths. If the scientific goal were to analyze strong SZ sources, then it may be more appropriate to only consider the error and bias of the strong signals, while the opposite would be true for weak signals.

\subsection{SZU vs. NILC}
The SZU model was selected by rank-ordering the total error and bias for all three classes of signal strength which was the $\tilde{y}_{0}=60$ model. This yielded a relatively unbiased model that was very comparable to NILC. Their similarity is visually apparent in \autoref{fig:example_plot} and statistically shown in \autoref{fig:pixel_compare} when evaluating the comparison sample. SZU yielded slightly larger MSEs for most of the SZ signal strengths compared to NILC, but the differences were not more than a few percent. Both NILC and SZU were dominated by the statistical errors with only a marginal contribution from the bias. If we carefully fine-tuned SZU and NILC, we could likely reach even more similar extractions.

The performances of SZU and NILC were also comparable when inspecting PSZ2 objects \autoref{fig:YSZ_compare}. Their extracted Y$_{\mathrm{SZ}}$ values closely followed a 1:1 relation, and there were no striking changes with Galactic latitude. One might expect Galactic latitude to play a role if the two methods handle foreground contamination in different ways. Interestingly, this trend was not present which may suggest that SZU and NILC inherently capture similar features.

\subsection{Angular Overlap Between Train/Test Data}
One potential concern is the construction of the training and test datasets, particularly due to the angular overlap between them. Both datasets include SZ signals injected into the same real-sky background, which may raise questions about their independence. Specifically, one might worry that the model is overfitting to the training data\textemdash effectively memorizing the background structure and performing background subtraction rather than learning the SZ signal. From this perspective, it could be argued that the test dataset should be angularly separated from the training set to eliminate any such dependence. However, we contend that this is unnecessary and that the current setup remains sufficient.

To clarify, consider an unrealistic, but idealized dataset where we could simulate a vast amount of SZ maps and we have Planck frequency maps where there are no known SZ signals, so the background contains pure contamination. Under this setup, we could inject all strengths of SZ signals without having to worry about masking. The model would see the full range of signals across all locations of the sky, and it would learn to separate the contamination and SZ components cleanly. Then we could take this model and apply it to the real sky to get a very accurate prediction. This is the conceptual framework our approach attempted to approximate.

If this scenario were possible, then the only difference between the training and test sets would be the different injected SZ maps. There is an infinite way to distribute the SZ signals, so there can always exist a test set that is different from the training data. Even when there is spatial overlap between the training and test data, the amplitudes, distributions, and morphologies of the SZ signals are always distinct between the training and test data. In other words, while the sky is fixed, the signal is not—each training and test example contains a different realization of the SZ component, drawn from a large and effectively uncorrelated set. The model must therefore learn to distinguish signal from structured noise across a wide dynamic range, which requires generalization rather than simple subtraction. 

Unfortunately, this perfect setup is impossible since we cannot obtain realistic frequency maps without an underlying SZ background. To overcome this, we injected only strong signals ($\tilde y >10$) and avoided regions of known galaxy clusters. Indeed, the masked regions of known clusters are where predictions would be most useful for science. We assume the contamination properties present in the masked regions are similar to their ambient regions that were available during training. 

Introducing strict angular separation between the training and test sets would undermine this key assumption. For example, if we trained the model exclusively on the northern Galactic hemisphere and tested on the southern hemisphere, the model would not encounter the full range of spatial variability present in the noise. As a result, we would expect degraded performance in regions it has not seen before i.e., the model has not seen the full variability in contamination. Our goal, instead, is to leverage as much of the sky as possible in both training and testing in order to expose the model to the diverse contamination patterns and enable more robust signal separation.

\subsection{Future Improvements}
There are some areas where there could be immediate improvements for machine learning extractions of the SZ signal. First, the square cutouts used for training/testing were $\sim 4^{\circ} \times 4^{\circ}$ which set a maximum angular size for this study. In the future, training data should be developed in order to capture SZ signals on larger angular scales by using larger patch sizes. Eventually, a method should be developed such that it can be applied to the entire sky.

Following that note, transformer-based architectures offer a potential avenue for capturing more global contexts \citep{Vaswani17}. These implement positional embeddings to help networks connect semantic information from different parts of an image. This may be beneficial since the intensity of foreground components change with position on the sky. Incorporating such information could help the network adapt to changes in Galactic latitude which is strongly correlated with foreground contamination. Instead of considering each cutout as an individual image, transformers would consider them as patches of a larger entity. Other architectures should also be explored in the fast-paced machine learning field, such as leveraging the advent of diffusion models. These have proven to be excellent at extracting underlying data distributions to solve a multitude of tasks such as removing structured noise like foreground contamination \citep{Stevens23}. 

In addition, an independent validation set of simulated data should be introduced in addition to that of \citet{Han21}. We believe the data used in this work were gathered in the best way currently possible, however, they are not perfect representations of the real Universe. For example, the synthetic SZ signals did not include signals from local systems such as nearby galaxies and galaxy groups. 

{More importantly, the simulated SZ signals in this work did not incorporate the well-established spectral and spatial correlations with the cosmic infrared background (CIB) largely caused by the emission from thermal dust in cluster member galaxies \citep{PlanckSZCIB,McCarthy23}. The CIB is known to generate spurious SZ-like signals, and this is particularly important for unresolved sources in Planck data, which includes a large number of galaxy clusters. As the CIB constitutes a significant source of contamination, its exclusion represents a limitation of the present analysis and should be addressed in future work that use simulation-based models.

Lastly, we want to emphasize the most significant advancements will be made from new observations with higher spectroscopic sampling. In some sense, it was not surprising that SZU and NILC yielded similar extractions since they utilized the exact same frequency data. The biggest challenge when separating the SZ signal from other components is caused by the varying emission laws. More spectral coverage will provide essential information regarding the amount of presence of each particular component.

\section{Summary} \label{sec:summary}
The main objective of this work was to build a machine learning model to extract the thermal SZ effect. To do this, we injected synthetic SZ signals from \citet{Han21} into real frequency observations from the \Planck satellite. We could have simulated the foregrounds as well and applied domain adaptation techniques, but this would have introduced additional challenges which we describe in \autoref{sec:domain_adapt}. Then we trained a three-dimensional convolutional neural network end-to-end with supervised learning to extract the SZ signals. 

When training SZU, we selected a subset of pixels by placing a lower limit on signal strength. This was a necessary step due to the superposition of simulated SZ signals being injected to the real sky. We set a floor of $\tilde{y}_{\mathrm{sig}}\geq 10$ to allow for a sizable sample while keeping the noise level in the labels manageable. Furthermore, we implemented curriculum learning which exposed the model to strong SZ signals at first and gradually introduced weaker ones. This depended on the starting signal amplitude $\tilde{y}_{0}$ which we set to 10 or 60 where the former denoted the absence of curriculum learning.

During evaluation, we considered three classes of signal strengths: weak, intermediate, and strong. We found that no curriculum learning yielded smaller overall error and bias for weaker signals. Including curriculum learning returned more unbiased solutions, and returned lower overall errors for the intermediate and strong signals.

We selected an SZU model for prediction using a rank-ordering method of the validation errors and biases among the three signals strengths. Then we compared the performance of SZU with the well-known NILC algorithm. We selected the curriculum learning model and found comparable results with NILC when evaluating simulated data. Then we extracted Y$_{\mathrm{SZ}}$ values from the PSZ2 catalog and found that SZU and NILC closely follow a 1:1 relation.

We concluded by mentioning possible avenues for improvement. Given the current state of SZ extraction methods, the choice of hyperparameters dictate the amount of bias-variance tradeoff. This must be taken into consideration for different scientific investigations, especially when stacking weak signals. Studies should anticipate the SNR in advance to help select an optimal extraction method. In addition, different model architectures and computer vision techniques should be explored to improve SZ feature extractions. Finally, we highlight that higher spectral resolution measurements will be the biggest factor in improving the component separation task.

%We would like to thank the anonymous referee for providing excellent feedback to help improve this work. 

We would like to thank the anonymous referee for helping to improve this work. We are also grateful for support from the NASA award 80NSSC22K0481/21-ADAP21-0069. This project made use of the Great Lakes computing cluster at the University of Michigan\textendash Ann Arbor. Code for this project is publicly available \citep{SZUCode}.

%In support of reproducible results, we provide code for this project at \url{https://github.com/campratt/multispectral/sz}.

%%%%%%%%%%%%%%%%%%%%%%%%%%%%%%%%%%%%%%%%%%%%%%%%%%%%%%%%%%%%%%%%%
\clearpage

%\bibliography{bib.bib}
%\bibliography{main.bib}{}

\bibliographystyle{aasjournal}

\appendix
\section{Domain Adaptation} \label{sec:domain_adapt}
In this work, we utilized synthetic SZ signals to train our models by superimposing them with the \Planck frequency data. The advantage of doing this was to expose the model to real-world contamination and adjust accordingly. Doing so, however, required that we place a lower limit on the strength of the SZ signal used in the loss function. This problem could be averted if we trained a model solely with synthetic data. This way we would not have to worry about the effects of superimposing simulated signals with an unknown background. Fully synthetic data, however, cannot capture the intricacies of the real sky. In turn, models trained with these data alone could introduce unwanted systematic effects during prediction. 

Attempting to adjust a synthetically trained model to real-world data could be achieved through domain adaptation. Domain adaptation transfers knowledge learned from a source domain to a target domain. In this context, the source domain would consist of labeled synthetic data while the target domain would be the real sky. In other words, the target domain is what would be used for scientific purposes. 

Domain adaptation may be a straight forward task in some machine learning applications. For example, one could train a model to recognize trees in a forest using data from a single forest (source domain), then apply the model to identify trees in an unseen forest (target domain). This works fine when the tasks are relatively simple, such as detecting trees, and the source and target domains are similar. This is not necessarily true when extracting the SZ signal. 

Extracting the SZ effect is a challenging task because it is a weak signal embedded in a sea of much stronger components. Also, full-sky simulations of the microwave sky are currently over-simplified. Simulating foreground emissions rely on strong assumptions, such as using a modified blackbody for the thermal dust emission law \citep{Thorne2017}. In turn, separating the SZ signal from a simulated dust component may not transfer easily to the real sky. Thus, the domain adaptation becomes complicated and may not lead to meaningful models.

Model selection would also become tricky in this sense of domain adaptation. As done in this work using a single domain, the best model is selected by evaluating its performance on an unseen, labeled, validation set. As previously mentioned, this does not exist for the SZ effect for real-world data, so there is no canonical model selection technique for this context. The question becomes: how does one select a model that performs well on the target data when the target data cannot be tested? This is an underappreciated problem that has only been recently explored in a few realms of machine learning. 

We considered using domain adaptation, however, we decided to not pursue this option. Rather than simulating foreground emissions, we decided to inject synthetic SZ signals into the real frequency observations. This has a couple of notable advantages. First, we do not have to make any assumptions about the underlying emission laws, so the models learn the foregrounds directly. Second, model selection is much easier since we can simply evaluate them on a labeled set of validation data since they exist in the same domain.



\end{document}